\documentclass[a4paper]{spie}  

\usepackage{amsmath,amsfonts,amssymb}
\usepackage{graphicx}
\usepackage{tikz}
\usetikzlibrary{positioning,arrows.meta,fit,shapes.geometric,backgrounds,calc}
\usepackage[colorlinks=true, allcolors=blue]{hyperref}

\newlength{\shotw}

\title{MIRA: a data management and education platform connecting students to robotic telescopes}

\author[a,b]{Peter P. Pedersen}
\author[c]{Lionel Garcia}
\author[d]{Mahshid Alimi}
\author[a]{David Degen}
\author[a]{Florian Lienhard}
\author[a,b]{Didier Queloz}

\affil[a]{ETH Z\"{u}rich, Department of Physics, Wolfgang-Pauli-Strasse 2, 8093 Zurich, Switzerland}
\affil[b]{Cavendish Laboratory, University of Cambridge, Cambridge, United Kingdom}
\affil[c]{Independent Technical Consultant, Perpignan, France}
\affil[d]{Independent Researcher, Z\"{u}rich, Switzerland}

\authorinfo{Further author information: send correspondence to P.P.P.\\P.P.P.: E-mail: ppedersen@ethz.ch}

\begin{document}
\maketitle

\begin{abstract}
Hands-on telescope experience is often used to drive student engagement in astronomy education, but scaling access to larger groups of students is operationally challenging. Consequently, students encounter only a fraction of the professional workflow, rarely engaging with the rigorous peer-review, time-allocation processes, or automated data reduction pipelines that govern modern research facilities.
We present the design of MIRA (Mentored Investigations using Robotic Astronomy), a data management and educational platform that connects Swiss secondary school and undergraduate students with operational robotic observatories. 
MIRA structures the entire observation lifecycle: proposal, review, acceptance/rejection, scheduling, and observation.
Following execution, the platform automatically reduces raw FITS frames (including astrometric calibration and photometry) and serves them via a web-accessible archive accompanied by Python-based analysis tutorials.
By separating educational front-ends from low-level telescope controls through Astra and ASCOM Alpaca, MIRA delivers an authentic scientific research workflow that bridges classroom learning with professional observatory operations.
\end{abstract}


\section{INTRODUCTION}
\label{sec:intro}

Autonomous robotic telescopes provide access to research-grade observations without requiring local instrumentation or specialist operational training.\cite{Gomez2017,Astra,debecker2026accessastronomicalobservationfacilities} Gomez and Fitzgerald distinguish remotely controlled telescopes from autonomous queued-request systems and identify the latter as the model that can serve larger educational communities.\cite{Gomez2017} The Las Cumbres Observatory global network applies this model at scale through its Global Sky Partners programme, which provides telescope time to education partners worldwide.\cite{Brown2013,Harbeck2024} Other educational services provide access to remote or autonomous observations, including the National Schools Observatory, Stellarium Gornergrat, and Skynet.\cite{Cooper2019,Gschwind2024,Zola2021,Reichart2023}

These platforms demonstrate the value of robotic observations, accessible request interfaces, teaching materials, and tools for image processing and scientific analysis in astronomy education.\cite{Cooper2019,Gschwind2024,Zola2021,Reichart2023} MIRA builds on these capabilities by combining automated telescope operations with mentor review. Its educational design gives students and teachers a representative workflow for planning and carrying out astronomical observations: students develop a scientific and technical case for an observation, mentors review the request, and the platform automates scheduling, execution, reduction, and archive delivery.

We present MIRA (Mentored Investigations using Robotic Astronomy), a platform for Swiss secondary-school students and ETH Z\"urich undergraduates. MIRA links proposal drafting, supervisor review, constraint-based scheduling, telescope execution, automated data reduction, archive access, and analysis tutorials within one workflow. By separating the educational interface, workflow engine, and telescope-control service, MIRA is designed to support mentor-guided learning while automating routine observatory operations for larger student cohorts.

This paper presents the MIRA platform design and evaluates its scheduling component on simulated observing nights. Section~\ref{sec:workflow} describes the proposal and observation workflow, including scheduling, automated reduction, and archive. Section~\ref{sec:architecture} describes the software architecture and component integration. Section~\ref{sec:summary} summarises the current validation and future work.

\section{PROPOSAL AND OBSERVATION WORKFLOW}
\label{sec:workflow}

Each observing proposal in MIRA is a \emph{plan}, which records the observing request (target, telescope, requested date, filter, and exposure time) together with the scientific and technical justifications. Each plan is in one of four review stages: draft, submitted, approved, or rejected. Approval triggers one or more observations that proceed through scheduling, execution, and archiving.

\subsection{Drafting and Submission}
\begin{figure}[ht]
\centering
\includegraphics[width=\columnwidth]{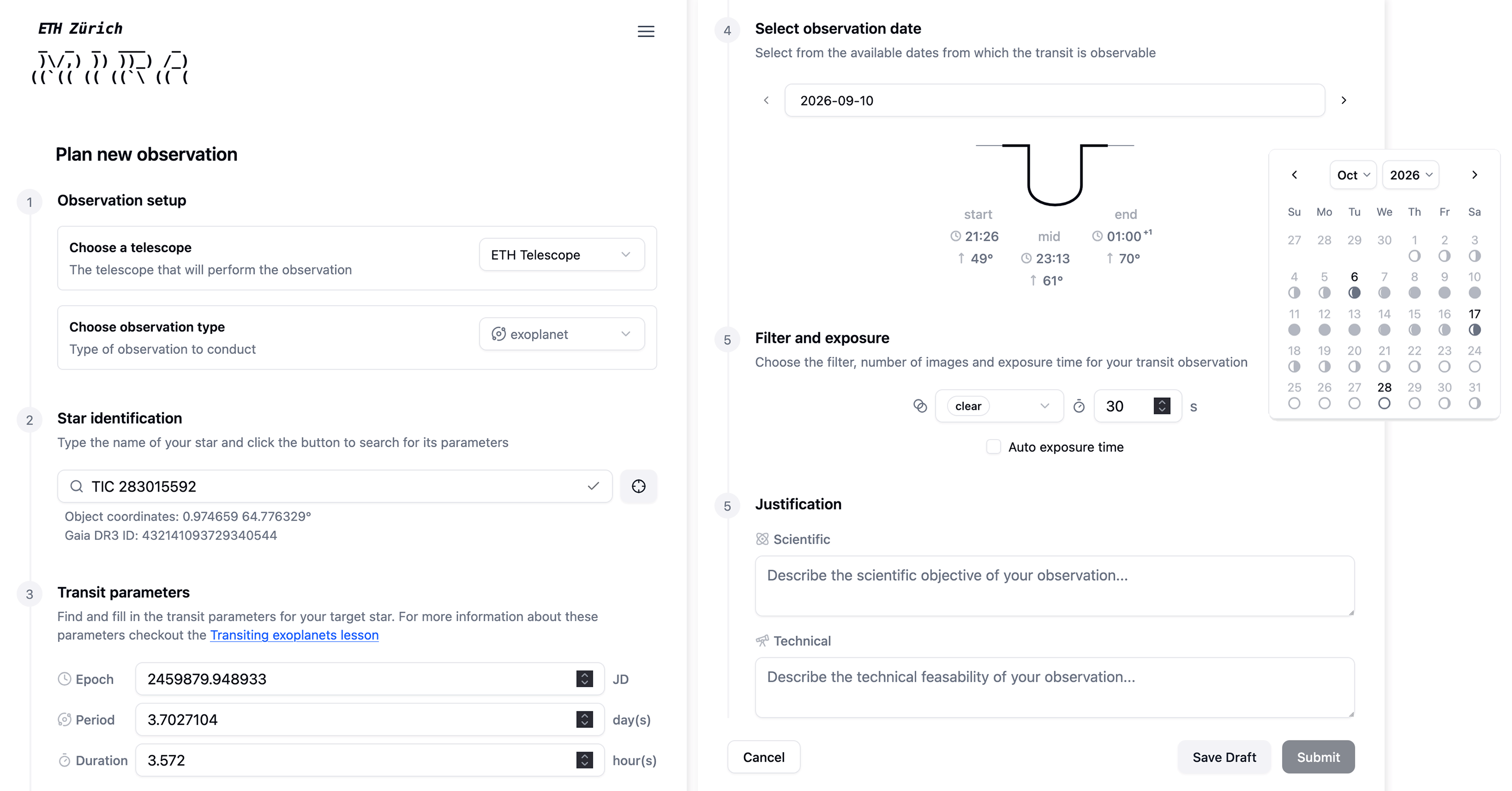}
\caption{The plan editor for an exoplanet-transit proposal. The student identifies the target star and its transit parameters, picks an observable night from the visibility calendar, sets the filter and exposure time, and enters the scientific and technical justification for review.}
\label{fig:plans}
\end{figure}

For the student workflows described here, a user opens the plan editor and selects either the imaging or exoplanet-transit template (Figure\,\ref{fig:plans}). When the user enters a target name, MIRA resolves its coordinates through SIMBAD\cite{Wenger2000} and displays a visibility calendar. Next, the user configures one or more filter blocks, each defined by an exposure time and a number of frames. The imaging template accepts several filters. The transit template restricts the request to a single filter without a defined number of frames, and additionally records the orbital period, transit epoch, and transit duration that the scheduler needs in order to place the observation on a transiting night. Finally, the student must provide both a scientific and technical justification for the observation before submitting the plan to the review queue.

\subsection{Review}
\begin{figure}[ht]
\centering
\includegraphics[width=0.8\columnwidth]{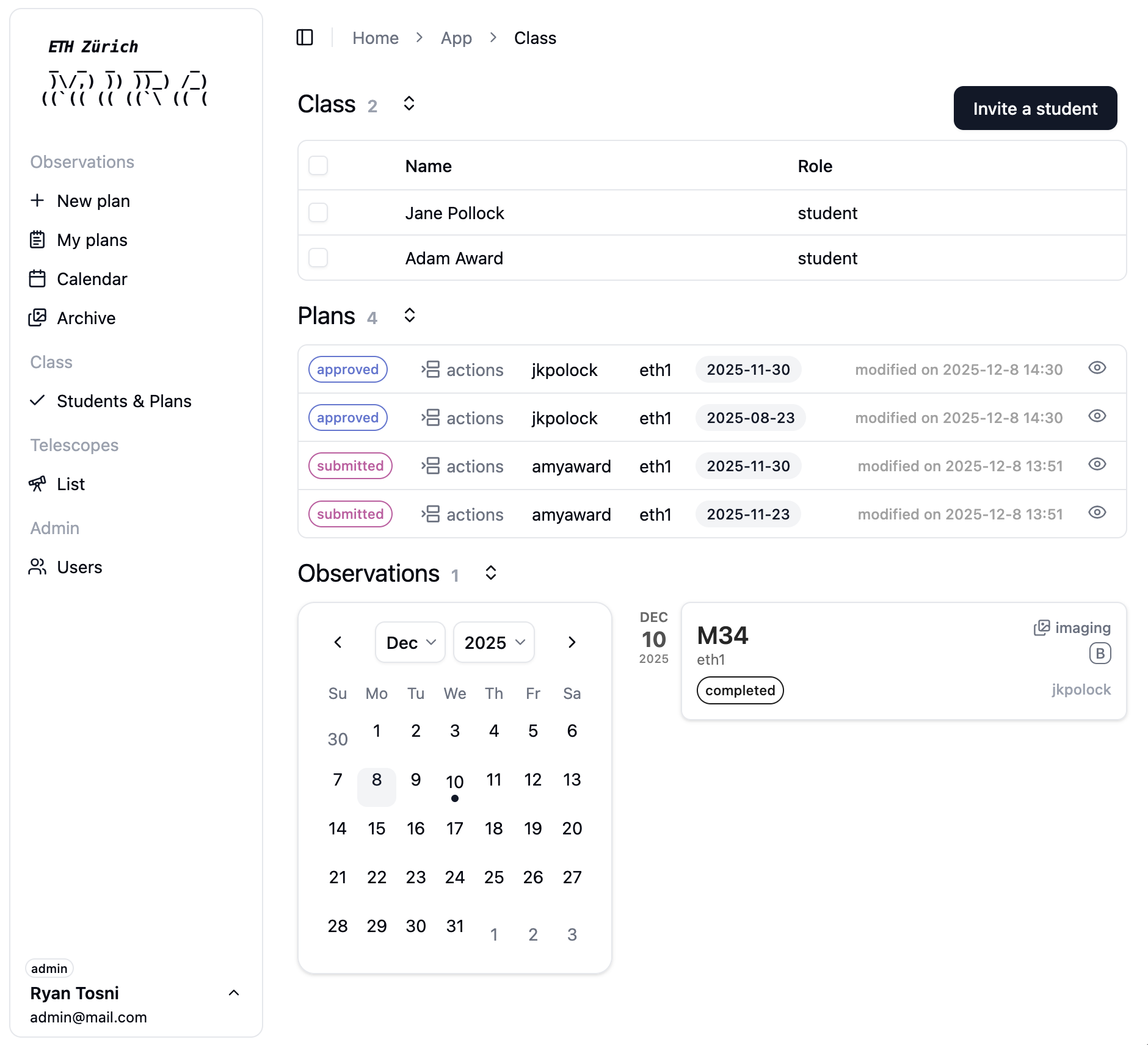}
\caption{The class view used by supervisors and administrators, listing enrolled members and their roles, the review status of each plan, and the observations that result from approved plans.}
\label{fig:admin}
\end{figure}

Using a centralised class view, a supervisor or administrator reviews the queued proposal, either approving the plan or returning it with feedback for revision (Figure\,\ref{fig:admin}). This dashboard lists each student's plans by status along with their active observations. Approval marks the transition from manual review to automated execution: it triggers scheduling, after which no human action is required until the data return.

\subsection{Scheduling}
\label{sec:scheduling}

This subsection first describes the scheduling algorithm and then evaluates it against three baseline planners on simulated observing nights.

\subsubsection{Scheduling Algorithm}

Once a proposal has been approved, MIRA converts its plan into an executable nightly schedule by evaluating physical feasibility at a discrete one-minute resolution. The scheduling engine treats observations as non-preemptive, which means that a scheduled item cannot begin until the previous one has finished. A candidate observation is feasible if its start and end times fall within the window specified in the proposal, the target stays above the local altitude limit for the full duration, and environmental conditions remain acceptable throughout, as determined by a per-minute timeline derived from weather forecasts. Time-critical observations, such as transits, carry the additional constraint that the transit centre must fall within the observation window. The scheduler therefore either places such an observation inside its permitted transit window or drops it from that night's schedule.

Among the feasible slots, the planner selects the sequence that maximises the total weighted score over the night. The reward function assigns a score to observation $j$ at candidate start time $t$:
\begin{equation}
\text{reward}(j, t) = w_{p}P_j + w_{\phi}\Phi_{j,\,t} - w_{a}A_{j,\,t} + w_{w}W_{j,\,t},
\label{eq:reward}
\end{equation}
where $P_j$ is the proposal priority assigned by MIRA, with higher priorities assigned to time-critical requests. $A_{j,\,t}$ is the mean airmass during the observation. $\Phi_{j,\,t}$ rewards observations that begin near the desired orbital phase, such as an exoplanet transit. $W_{j,\,t}$ represents additional weather-forecast information when available. The default weights in Equation~(\ref{eq:reward}) are $w_p=1$, $w_\phi=1$, $w_a=0.5$, and $w_w=0.35$. One can adjust these weights for different scheduling objectives.

MIRA finds the optimal schedule with an exact subset dynamic-programming formulation. The method evaluates combinations of the candidate observations and is suitable for the small candidate sets expected during one night. At minute $t$, the state contains $S$, the set of candidate observations that remain unscheduled. The state-value function $V(t, S)$ gives the maximum cumulative reward available from minute $t$ to the end of the night. The engine computes the values of this function by backward induction, from the final minute of the night to the present, using the Bellman recursion:%
\cite{Bellman1954}
\begin{equation}
V(t, S) = \max \left( Q_{\text{idle}}(t, S), \, \max_{j \in S} Q_{\text{run}}(j, t, S) \right).
\label{eq:bellman}
\end{equation}
The action-value functions $Q_{\text{idle}}(t, S)$ and $Q_{\text{run}}(j, t, S)$ represent the total projected payoffs of choosing to either idle or execute a specific observation $j$:
\begin{align}
Q_{\text{idle}}(t, S) &= -\text{idle\_cost} + \gamma V(t+1, S), \\
Q_{\text{run}}(j, t, S) &= \text{reward}(j, t) + \gamma^{\text{duration}(j)}
V\left(t + \text{duration}(j), \, S \setminus \{j\}\right).
\end{align}
The $\text{idle\_cost}$ term penalises unallocated time to minimise dead time during clear skies. The temporal discount factor $\gamma \in (0, 1]$ gives higher weight to earlier observations over later ones, mitigating the compounding uncertainty of weather forecasts. The default values, used in all simulations below, are $\text{idle\_cost} = 0.01$ and $\gamma = 0.999$. The parameter $\text{duration}(j)$ represents the total time allocated to observation $j$. $S \setminus \{j\}$ denotes the set difference, removing the completed observation $j$ from the candidate pool for all subsequent time steps. 
Solving this recursion yields the optimal schedule, which may leave some candidates unscheduled when no feasible or sufficiently valuable placement remains. This exact formulation scales as $\mathcal{O}(2^N)$ with the number of candidate observations $N$.

\subsubsection{Evaluation against Baseline Planners}

We compare the Bellman planner against two baselines from astroplan: the \texttt{PriorityScheduler}, which uses greedy priority scheduling, and the \texttt{SequentialScheduler}, which processes a fixed queue order. We additionally compare it with SCOPES, a beam-search heuristic planner.%
\cite{Morris2018,10.1117/12.2311839}
Each planner receives equivalent scheduling objectives and the same feasibility information, including the per-minute weather gate, the altitude limit, and the transit-coverage windows. The astroplan schedulers enforce transit windows through per-block time constraints and score placements with a continuous airmass constraint. SCOPES instead applies transit windows through a veto merit and uses a Gaussian timing merit, matched to the Bellman phase term, to favour centred transits.

We simulated 100 synthetic nights under identical conditions. Each simulation began at 19:00~UTC and spanned nine hours. A night contained 6--10 candidate observations with durations of 20--90 minutes. We placed targets near the meridian at random times. Time-sensitive transit observations accounted for 20\% of the candidates, and their epochs fell within the simulated night. We modelled the weather as a clear night containing 0--2 unsafe intervals of 30--120 minutes. Sixty simulated nights contained at least one unsafe interval.

We evaluated each planner using two measures. The first was the ratio between the planner's mean airmass, calculated over its own scheduled observations, and the Bellman planner's mean airmass for the same night; a ratio greater than one indicated that the planner observed through more atmosphere. Because this ratio is computed only over each planner's own schedule, it is sensitive to how many observations a planner completes: a planner that schedules fewer, easier-to-place observations can show a favourable ratio without necessarily placing shared observations more efficiently. We therefore report this measure alongside the number of observations scheduled per night, and interpret airmass differences with that count in mind. The second measure was transit capture, defined as the fraction of the 158 transit candidates scheduled with an observation that contained the transit centre. Every planner treated the transit-coverage window as a hard constraint, so this measure records how many transit observations each planner actually scheduled.

\begin{figure}[ht]
  \centering
  \includegraphics[width=\linewidth]{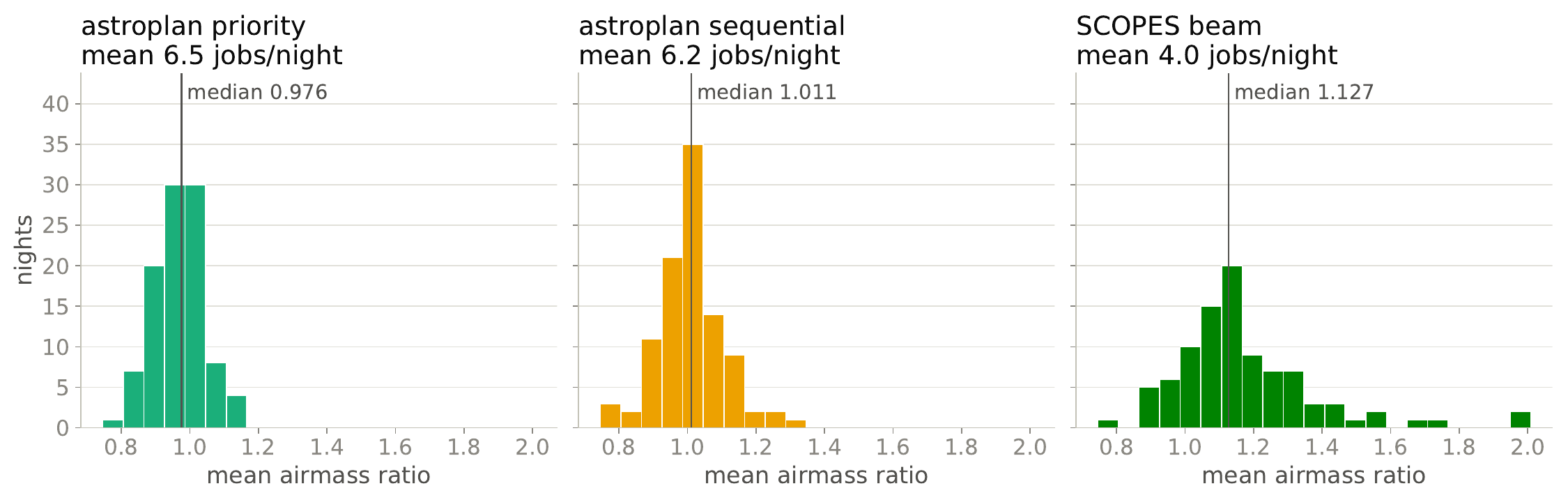}
  \caption{Per-night mean airmass of each baseline planner relative to the Bellman planner over 100 synthetic nights (9-hour horizon, $\le$10 candidates per night, 20\% time-sensitive transits). The ratio divides the baseline's mean airmass over its own scheduled observations by the Bellman planner's value for the same night; values above 1 indicate higher (worse) airmass. Solid vertical lines mark the median ratio. Panel headers give the mean number of observations scheduled per night; the Bellman planner averaged 7.0. The SCOPES panel excludes the 7 nights on which it returned no schedule.}
  \label{fig:airmass_hist}
\end{figure}

Figure\,\ref{fig:airmass_hist} shows the airmass-ratio distributions. The astroplan priority baseline achieved a median ratio of 0.98 and produced lower mean airmass than the Bellman planner on 68\% of nights. It scheduled 6.5 observations per night, compared with 7.0 for the Bellman planner. Its greedy placements left idle gaps between observations. The astroplan sequential baseline had a median ratio of 1.01 and exceeded parity on 54\% of nights. It scheduled 6.2 observations per night. SCOPES beam had a median ratio of 1.13 and exceeded parity on 84\% of nights. It scheduled 4.0 observations per night.

SCOPES packed observations contiguously and did not preserve idle intervals. Unsafe-weather intervals and transit windows therefore terminated some of its schedules early. SCOPES returned an empty schedule on 7 of the 100 nights, which Figure\,\ref{fig:airmass_hist} excludes; transit capture counts these nights as zero captures. Transit capture distinguished the planners more strongly. The Bellman planner scheduled 125 of the 158 transit candidates (79\%). astroplan priority scheduled 121 candidates (77\%). astroplan sequential scheduled 59 candidates (37\%), and SCOPES beam scheduled 26 candidates (16\%); excluding the empty nights instead would raise this fraction to 17\%. The Bellman planner scheduled the greatest number of observations and captured the most transits while maintaining mean airmass close to that of the priority baseline.

\begin{figure}[ht]
  \centering
  \includegraphics[width=\linewidth]{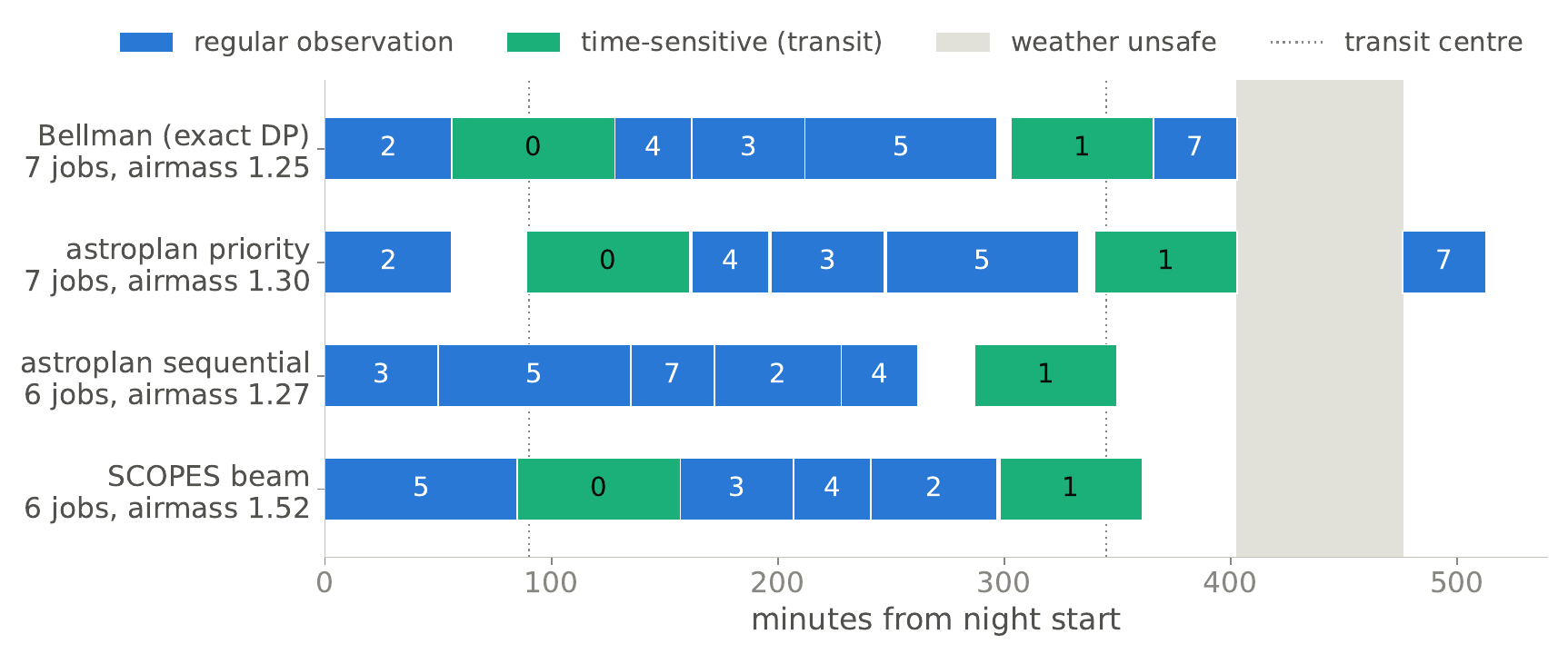}
  \caption{Schedules produced by the four planners for a single synthetic night (2026-12-11, 8 candidate observations). Each lane represents one planner; numbered bars denote observation intervals, coloured by type (blue: regular; green: time-sensitive transit). Dotted vertical lines indicate the transit centres, and the shaded band marks an unsafe-weather interval. Lane labels show the number of scheduled observations and the mean airmass. Every scheduled transit contains its centre, as the coverage window requires. The Bellman planner schedules seven observations without gaps and finishes before the weather block. astroplan priority defers observation~7 until after the block and observes it at high airmass; astroplan sequential omits one of the two transits.}
  \label{fig:example_night}
\end{figure}

Figure\,\ref{fig:example_night} shows the schedules for one simulated night. Each planner placed any scheduled transit within its required coverage window. The Bellman planner scheduled seven observations consecutively and completed them before the unsafe-weather interval. The astroplan priority baseline also scheduled seven observations. It left gaps around its airmass-optimal placements and deferred observation~7 until after the weather interval, when the target had high airmass. The astroplan sequential baseline omitted one of the two transit observations.

The reward weights in Equation~(\ref{eq:reward}) determine the trade-off between transit timing, airmass, and unused time. The current values represent initial assumptions about the composition of the observing queue. Operational use of MIRA will provide the data needed to refine these weights and to assess the balance between time-critical and flexible requests.

\subsection{Execution, Reduction, and Archiving}
\begin{figure}[ht]
\centering
\includegraphics[width=0.8\columnwidth]{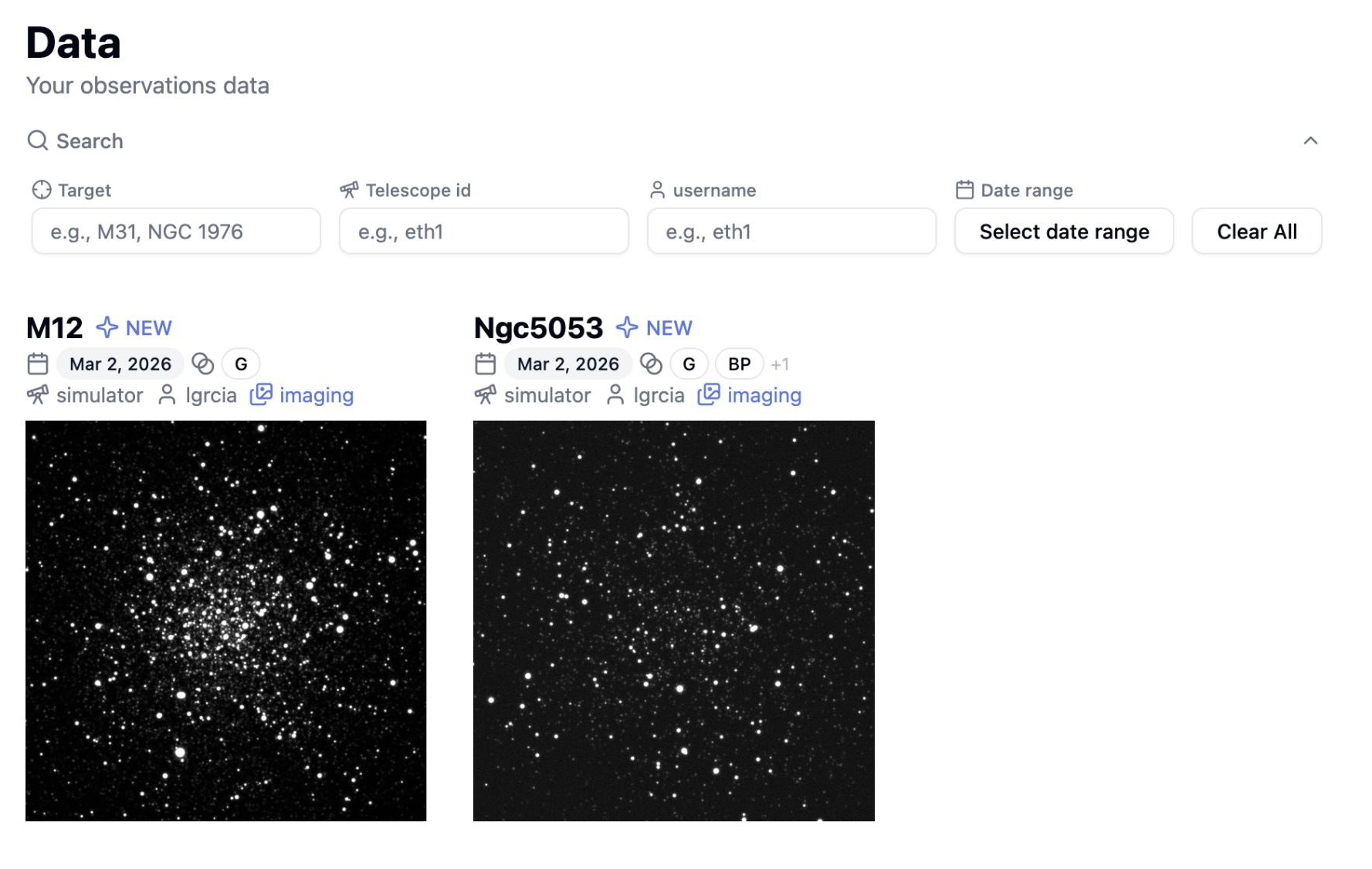}
\caption{The archive view, searchable by target, telescope, user, and date, showing the reduced image for each completed observation.}
\label{fig:archive}
\end{figure}

On the assigned night, the schedule is transmitted to the telescope-control service, which executes it and writes the FITS frames to the shared volume, reporting each observation through the API as it moves from \texttt{planned} to \texttt{observing} and then to \texttt{completed}, or to \texttt{failed} on error. Once the FITS frames are available at the end of a night, MIRA initiates automated data reduction: it builds master calibration frames, calibrates and aligns the science images, matches sources to the Gaia catalogue,\cite{GaiaDR2,Ginsburg2019} and performs aperture photometry. The resulting products and metadata are stored in the public archive, browsable by target, telescope, and date, and available for download (Figure\,\ref{fig:archive}). Alongside the archive, MIRA provides interactive Python/Jupyter tutorials that guide students through the same reduction and analysis steps. The automatic photometry pipeline can be delayed by the teacher to allow students to analyse first before comparing to the reference pipeline.

\section{SYSTEM ARCHITECTURE}
\label{sec:architecture}

\begin{figure}[t]
\centering
\begin{tikzpicture}[
  font=\footnotesize, >=Latex,
  box/.style={draw, rounded corners, align=center, minimum height=9mm, minimum width=28mm, fill=white, inner sep=2pt},
  store/.style={draw, rounded corners, align=center, minimum height=8mm, fill=black!4},
  grp/.style={draw, dashed, rounded corners, inner sep=3mm},
  ttl/.style={font=\footnotesize\bfseries, anchor=south west},
  lbl/.style={font=\scriptsize, fill=white, inner sep=1.5pt}
]
\node[box, minimum width=20mm] (browser) at (-3.2,0) {Browser};
\node[box] (client) at (0,0) {Web application\\(SvelteKit + API)};
\node[box] (sqlite) at (0,-1.7) {Metadata store\\(SQLite / Drizzle)};
\node[box, minimum height=12mm, text width=30mm] (airflow) at (5,0) {Workflow engine\\(Airflow + Celery worker)};
\node[box] (astra) at (10,0) {Telescope control\\(Astra / Alpaca)};
\node[box] (hw) at (10,-1.7) {Telescope\\/ simulator};
\node[store, minimum width=130mm] (vol) at (5,-3.6) {Shared data volume \;\;(schedules, FITS images)};

\begin{scope}[on background layer]
  \node[grp, fit=(client)(sqlite)] (gA) {};
  \node[grp, fit=(airflow)] (gB) {};
  \node[grp, fit=(astra)(hw)] (gC) {};
\end{scope}
\node[ttl] at (gA.north west) {Interface};
\node[ttl] at (gB.north west) {Workflow};
\node[ttl] at (gC.north west) {Acquisition};

\draw[->] (browser) -- (client) node[lbl, midway, above]{HTTP};
\draw[<->] (client) -- (sqlite) node[lbl, midway, right]{SQL};
\draw[->] (client) -- (airflow) node[lbl, midway, above]{trigger DAG};
\draw[->] (astra) -- (hw) node[lbl, midway, right]{Alpaca};

\draw[->] (vol.north -| sqlite) -- (sqlite.south) node[lbl, midway, right]{archive reads};
\draw[->] (airflow.south) -- (airflow.south |- vol.north) node[lbl, pos=0.55, right]{schedule};
\draw[->] (hw.south) -- (hw.south |- vol.north) node[lbl, midway, right]{FITS};

\draw[->] (astra.north) -- ++(0,1.5) -| (client.north) node[lbl, pos=0.25, above]{status};
\end{tikzpicture}
\caption{\label{fig:architecture} Implementation architecture of MIRA. The interface layer manages plans, users, and archive access. The workflow layer performs scheduling and reduction. The acquisition layer executes schedules on a telescope or simulator. The implementation names appear in parentheses.}
\end{figure}
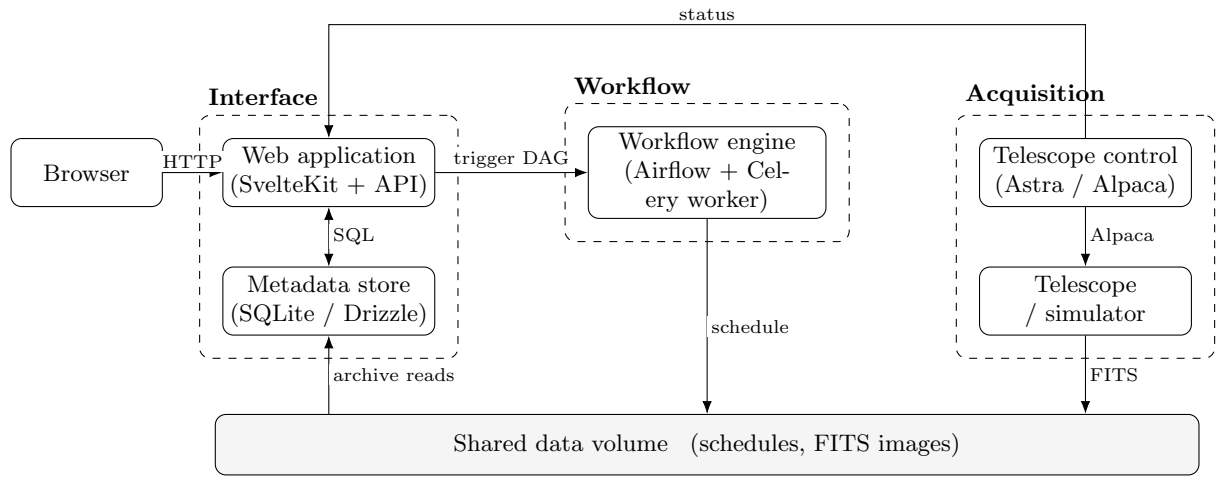

MIRA is organised as three independent, containerised components: a web application, a workflow engine, and a telescope-control service. Table\,\ref{tab:components} summarises their roles, and Figure\,\ref{fig:architecture} illustrates how they exchange schedules, metadata, and FITS files.

\begin{table}[t]
\centering
\caption{Functional view of the main software components. The implementation names appear in the rightmost column.}
\label{tab:components}
\begin{tabular}{p{30mm}p{58mm}p{52mm}}
\hline
Function & Role in MIRA & Current implementation \\
\hline
Web application & Plan drafting, review, authentication, archive browsing, API endpoints & SvelteKit, SQLite, Drizzle \\
Workflow engine & Schedule search, schedule transfer, data reduction, derived products & Apache Airflow, Celery \\
Telescope control & Schedule execution, hardware communication, status reporting, simulator control & Astra over ASCOM Alpaca \\
\hline
\end{tabular}
\end{table}

The web application, built with SvelteKit\footnote{\url{https://svelte.dev/}},handles plan drafting, proposal review, authentication, and archive access, providing role-based permissions for students, supervisors, administrators, and observers.  Metadata is managed through the Drizzle object-relational mapper. While SQLite is currently used for simplicity, Drizzle supports alternative database engines such as PostgreSQL and MySQL.

The workflow engine uses Apache Airflow\footnote{\url{https://airflow.apache.org/}} to orchestrate scheduling and reduction tasks as Directed Acyclic Graphs (DAGs). A DAG breaks a pipeline into tasks with explicit dependencies, so each step can be retried, logged, and inspected independently without rerunning the entire pipeline. These tasks run on a Celery worker, which executes them asynchronously in the background to prevent long scheduling or reduction processes from blocking the web application.

The telescope-control service is Astra, a pre-existing observatory control software\cite{Astra, pedersen2026astra}. MIRA integrates with Astra through a custom observatory subclass that handles MIRA-specific operations, such as exposure logging and status reporting. To execute a schedule, the workflow engine writes a plan to the shared data volume, where Astra reads it directly. As the telescope executes the observations, Astra reports status updates and triggers automated data reduction by calling back into the web application API. This workflow decouples educational logic from low-level hardware control.

\section{SUMMARY AND FUTURE WORK}
\label{sec:summary}

MIRA provides an educational workflow that reflects the main stages of a professional observing programme. 
Students can prepare observing proposals, receive feedback from their supervisors, have their approved observations autonomously scheduled and executed, and access calibrated data products through a searchable archive. Additionally, MIRA offers Python/Jupyter tutorials that link the archived observations to scientific analysis.

The scheduling engine combines proposal constraints, target altitude, weather forecasts, and transit ephemerides in an exact dynamic-programming search. In simulations of 100 synthetic nights, the Bellman planner scheduled the largest number of observations and captured the largest fraction of time-critical transit requests. It maintained mean airmass close to that achieved by the greedy astroplan priority scheduler. These results support its use for the small nightly candidate sets expected in the initial MIRA deployment.

MIRA separates the web application, workflow engine, and telescope-control service into independent components. The workflow engine exchanges schedules and FITS products with Astra through a shared data volume, while the web application manages proposal review, user access, observation status, and archive browsing. This design separates educational workflows from low-level observatory control and supports simulator-based validation and planned deployment with physical instruments.

We validated the current implementation with an ASCOM Alpaca observatory simulator that generates synthetic images.%
\footnote{\url{https://github.com/ppp-one/alpaca-simulators}}
In the next phase, we will deploy MIRA on the 0.5\,m telescope at ETH Z\"{u}rich. This deployment will test the complete workflow under real observing conditions, including weather interruptions, hardware faults, calibration quality, data-transfer reliability, and telescope-control responses.

Future work will evaluate the platform with secondary-school and undergraduate users. These studies will assess whether proposal drafting, review, scheduling feedback, and analysis tutorials improve students' understanding of observational astronomy and research practice. Operational data will also guide refinement of the scheduling weights, expansion of the available observing templates, and development of monitoring tools for teachers and observatory staff.

\acknowledgments
The authors deeply thank members of Zurich University of the Arts (ZHdK) and Anna Vasvari for aiding in the design evolution of MIRA.

\bibliography{report}

@article{Gomez2017,
  author = {E. L. Gomez and M. T. Fitzgerald},
  title = {Robotic telescopes in education},
  journal = {Astronomical Review},
  volume = {13},
  number = {1},
  pages = {28--68},
  year = {2017},
  doi = {10.1080/21672857.2017.1303264},
  publisher = {Informa UK Limited},
}

@article{Gschwind2024,
  author = {Gschwind, S. and Hohmann, Sascha and Müller, André and Nordine, J. and Riesen, T.},
  title = {The {S}tellarium {G}ornergrat: astrophysics with your own data},
  journal = {Journal of Physics: Conference Series},
  volume = {2727},
  pages = {012011},
  year = {2024},
  number = {1},
  doi = {10.1088/1742-6596/2727/1/012011},
  publisher = {IOP Publishing},
}

@inproceedings{Harbeck2024,
  author = {Harbeck, Daniel R. and Taylor, Brook and Kirby, A. and Bowman, M. and Foale, S. and Kadlec, Kal and McCully, C. and Daily, M. and Vera, J. de and Douglas, Dave and Willis, Mark and Baker, Ian D. and Volgenau, N. and Conway, P. and Haworth, B. and Estrada, Jesus and Gomez, Edward C. and Seale, Sandy and Hopkinson, Alice and Rios, Fernando and Kotapali, Prerana and Storrie-Lombardi, L. and Rosing, W.},
  title = {An upgraded 0.4-meter telescope fleet for {L}as {C}umbres {O}bservatory's educational and science programs},
  booktitle = {Astronomical Telescopes + Instrumentation},
  editor = {J. J. Bryant and K. Motohara and J. R. D. Vernet},
  series = {Proc. SPIE},
  volume = {13096},
  pages = {131},
  year = {2024},
  journal = {Astronomical Telescopes + Instrumentation},
  doi = {10.1117/12.3014719},
  publisher = {SPIE},
}

@article{Brown2013,
  author = {Brown, T. M. and Baliber, N. and Bianco, F. B. and Bowman, M. and Burleson, B. and Conway, P. and Crellin, M. and Depagne, {\'E}. and De Vera, J. and Dilday, B. and Dragomir, D. and Dubberley, M. and Eastman, J. D. and Elphick, M. and Falarski, M. and Foale, S. and Ford, M. and Fulton, B. J. and Garza, J. and Gomez, E. L. and Graham, M. and Greene, R. and Haldeman, B. and Hawkins, E. and Haworth, B. and Haynes, R. and Hidas, M. and Hjelstrom, A. E. and Howell, D. A. and Hygelund, J. and Lister, T. A. and Lobdill, R. and Martinez, J. and Mullins, D. S. and Norbury, M. and Parrent, J. and Paulson, R. and Petry, D. L. and Pickles, A. and Posner, V. and Rosing, W. E. and Ross, R. and Sand, D. J. and Saunders, E. S. and Shobbrook, J. and Shporer, A. and Street, R. A. and Thomas, D. and Tsapras, Y. and Tufts, J. R. and Valenti, S. and Vander Horst, K. and Walker, Z. and White, G. and Willis, M.},
  title = {Las {C}umbres {O}bservatory global telescope network},
  journal = {Publications of the Astronomical Society of the Pacific},
  volume = {125},
  number = {931},
  pages = {1031--1055},
  year = {2013},
  doi = {10.1086/673168},
  publisher = {IOP Publishing},
}

@article{Morris2018,
  title = {astroplan: An Open Source Observation Planning Package in Python},
  volume = {155},
  ISSN = {1538-3881},
  url = {http://dx.doi.org/10.3847/1538-3881/aaa47e},
  DOI = {10.3847/1538-3881/aaa47e},
  number = {3},
  journal = {The Astronomical Journal},
  publisher = {American Astronomical Society},
  author = {Morris,  Brett M. and Tollerud,  Erik and Sipőcz,  Brigitta and Deil,  Christoph and Douglas,  Stephanie T. and Medina,  Jazmin Berlanga and Vyhmeister,  Karl and Smith,  Toby R. and Littlefair,  Stuart and Price-Whelan,  Adrian M. and Gee,  Wilfred T. and Jeschke,  Eric},
  year = {2018},
  month = Feb,
  pages = {128}
}

@article{GaiaDR2,
  author = {{Gaia Collaboration}},
  title = {Gaia {D}ata {R}elease 2: summary of the contents and survey properties},
  journal = {Astronomy \& Astrophysics},
  volume = {616},
  pages = {A1},
  year = {2018},
  doi = {10.1051/0004-6361/201833051},
}

@article{Wenger2000,
  author = {Wenger, M. and Ochsenbein, F. and Egret, D. and Dubois, P. and Bonnarel, F. and Borde, S. and Genova, F. and Jasniewicz, G. and Laloe, S. and Lesteven, S. and Monier, R.},
  title = {The {SIMBAD} astronomical database: the {CDS} reference database for astronomical objects},
  journal = {Astronomy \& Astrophysics Supplement Series},
  volume = {143},
  pages = {9--22},
  year = {2000},
  doi = {10.1051/aas:2000332},
}

@article{Ginsburg2019,
  author = {Ginsburg, A. and SipHocz, B. and Brasseur, C. and Cowperthwaite, P. and Craig, M. and Deil, C. and Guillochon, J. and Guzman, Giannina and Liedtke, Simon and Lim, P. and Lockhart, K. and Mommert, Michael and Morris, B. and Norman, H. and Parikh, Madhura and Persson, Magnus and Robitaille, T. and Segovia, J. and Singer, L. and Tollerud, E. and Val-Borro, Miguel de and Valtchanov, I. and Woillez, J. and collaboration, the Astroquery},
  title = {astroquery: an astronomical web-querying package in {P}ython},
  journal = {Astronomical Journal},
  volume = {157},
  number = {3},
  pages = {98},
  year = {2019},
  doi = {10.3847/1538-3881/aafc33},
  publisher = {American Astronomical Society},
}

@misc{Astra,
  doi = {10.5281/ZENODO.18890151},
  url = {https://zenodo.org/doi/10.5281/zenodo.18890151},
  author = {Pedersen,  Peter P. and Degen,  David and Garcia,  Lionel and Zúñiga-Fernández,  Sebastián and Sebastian,  Daniel and Schroffenegger,  Urs and Queloz,  Didier},
  title = {Astra},
  publisher = {Zenodo},
  year = {2026},
  copyright = {GNU General Public License v3.0 only},
}

@misc{debecker2026accessastronomicalobservationfacilities,
      title={How to access astronomical observation facilities ?}, 
      author={Michaël De Becker},
      year={2026},
      eprint={2606.26996},
      archivePrefix={arXiv},
      primaryClass={astro-ph.IM},
      url={https://arxiv.org/abs/2606.26996}, 
}

@inproceedings{10.1117/12.2311839,
author = {Ruby van Rooyen and Deneys S. Maartens and Peter Martinez},
title = {{Autonomous observation scheduling in astronomy}},
volume = {10704},
booktitle = {Observatory Operations: Strategies, Processes, and Systems VII},
editor = {Alison B. Peck and Robert L. Seaman and Chris R. Benn},
organization = {International Society for Optics and Photonics},
publisher = {SPIE},
pages = {1070410},
year = {2018},
doi = {10.1117/12.2311839},
URL = {https://doi.org/10.1117/12.2311839}
}

@article{Bellman1954,
  title = {The theory of dynamic programming},
  volume = {60},
  ISSN = {0273-0979},
  url = {http://dx.doi.org/10.1090/S0002-9904-1954-09848-8},
  DOI = {10.1090/s0002-9904-1954-09848-8},
  number = {6},
  journal = {Bulletin of the American Mathematical Society},
  publisher = {American Mathematical Society (AMS)},
  author = {Bellman,  Richard},
  year = {1954},
  pages = {503–515}
}

@article{Cooper2019,
  title = {The National Schools’ Observatory: Access to the Universe for All},
  volume = {15},
  ISSN = {1743-9221},
  url = {http://dx.doi.org/10.1017/S1743921321001186},
  DOI = {10.1017/s1743921321001186},
  number = {S367},
  journal = {Proceedings of the International Astronomical Union},
  publisher = {Cambridge University Press (CUP)},
  author = {Cooper,  Sally E.},
  year = {2019},
  month = Dec,
  pages = {28–29}
}

@article{Zola2021,
  title = {LONG-TERM PHOTOMETRY WITH SKYNET ROBOTIC TELESCOPE NETWORK},
  volume = {53},
  ISSN = {1405-2059},
  url = {http://dx.doi.org/10.22201/ia.14052059p.2021.53.40},
  DOI = {10.22201/ia.14052059p.2021.53.40},
  journal = {Revista Mexicana de Astronomía y Astrofísica Serie de Conferencias},
  publisher = {Universidad Nacional Autonoma de Mexico},
  author = {Zola,  S. and Kouprianov,  V. and Reichart,  D. E. and Bhatta,  G. and Caton,  D. B.},
  year = {2021},
  month = Sep,
  pages = {206–214}
}

@misc{Reichart2023,
  doi = {10.48550/ARXIV.2304.02545},
  url = {https://arxiv.org/abs/2304.02545},
  author = {Reichart,  Daniel E. and Haislip,  Joshua and Kouprianov,  Vladimir and Fu,  Ruide and Selph,  Logan and Xu,  Shengjie and Torian,  John and Keohane,  Jonathan and Janzen,  Daryl and Moffett,  David and Converse,  Stanley},
  title = {Next-Level,  Robotic Telescope-Based Observing Experiences to Boost STEM Enrollments and Majors on a National Scale: Year 1 Report},
  publisher = {arXiv},
  year = {2023},
  copyright = {Creative Commons Attribution Non Commercial Share Alike 4.0 International}
}

@inproceedings{pedersen2026astra,
  author    = {Pedersen, Peter P. and Degen, David and Garcia, Lionel and
               Schroffenegger, Urs and Sebastian, Daniel and
               Z{\'u}{\~n}iga-Fern{\'a}ndez, Sebasti{\'a}n and
               Demory, Brice-Olivier and Ducrot, Elsa and
               Gillon, Micha{\"e}l and Hooton, Matthew J. and
               Jan{\'o}-Mu{\~n}oz, Cl{\`a}udia and McCormac, James and
               Timmermans, Mathilde and Triaud, Amaury H. M. J. and
               Queloz, Didier},
  title     = {Astra: an open-source fully autonomous robotic observatory control software},
  booktitle = {Software and Cyberinfrastructure for Astronomy IX},
  series    = {Proc. SPIE},
  volume    = {14155},
  year      = {2026},
  note      = {in press, paper 14155-116}
}
\bibliographystyle{spiebib}

\end{document}